 \documentclass[final,5p,times,twocolumn]{elsarticle}

\usepackage{amssymb}
\usepackage{lipsum}
\usepackage{amsmath}

\usepackage{xcolor}
\usepackage{color}
\usepackage{float} 
\usepackage{graphicx}
\usepackage{soul}
\usepackage[T1]{fontenc} 
\setcitestyle{square} 
\biboptions{comma}
\usepackage[colorlinks=true, linkcolor=red, citecolor=green, urlcolor=magenta]{hyperref}

\begin{document}

\hypersetup{
    colorlinks=true,
    linkcolor=red,    
    citecolor=blue,   
    urlcolor=black      
}

\begin{frontmatter}



\title{Observational constraints on entropic cosmology}

\author[First]{Javier Chagoya}
\ead{javier.chagoya@fisica.uaz.edu.mx}
\author[First]{I. D\'iaz-Salda\~na}
\ead{isaacdiaz@fisica.uaz.edu.mx}
\author[First]{Mario H. Amante}
\ead{mario.herrera@fisica.uaz.edu.mx}
\author[First]{J. C. L\'opez-Domínguez}
\ead{jlopez@fisica.uaz.edu.mx}
\author[Second]{M. Sabido}
\ead{msabido@fisica.ugto.mx}
\affiliation[First]{organization={Unidad Acad\'emica de F\'isica, Universidad Aut\'onoma de Zacatecas},
addressline={Calzada Solidaridad esquina con Paseo a la Bufa S/N}, city={Zacatecas},
            postcode={98060}, 
            state={Zacatecas},
            country={ M\'exico}}
\affiliation[Second]{organization={Departamento de F\'sica de la Universidad de Guanajuato},
addressline={A.P. E-143}, city={Le\'on},    
            postcode={37150},
            state={Guanajuato},
            country={M\'exico}}    
\begin{abstract}
In this  work, we derive a generalized modified Friedmann equation based on an entropy-area relation that incorporates established modifications, such as volumetric, linear, and logarithmic terms, in addition to novel entropic modifications that might yield to relevant cosmological implications at different stages of the evolution of the Universe. Some of these modifications are capable of mimicking the effects of dark energy and describing the current state of accelerated expansion of the Universe.  We study particular cases of the generalized Friedmann equation and constrain the free parameters using observational datasets, including Hubble parameter measurements, baryon acoustic oscillations, and strong lensing systems. Our findings indicate that the proposed models align well with current observational data, particularly in low-redshift regimes; furthermore, these models are compatible with the value of $H_0$ obtained by the SH0ES program.
\end{abstract}


\end{frontmatter}
\section{Introduction}\label{introduction}
One of the most outstanding achievements in theoretical physics is the formulation of General Relativity (GR). The current observations in black holes and gravitational waves cement GR as the theory to describe the gravitational interaction. Although the open problem of dark energy and dark matter can be compatible with GR by proposing exotic sources of matter and energy, current observations do not discard alternative theories of gravity. Moreover, on a more theoretical front, the lack of a complete quantum theory of gravity after decades of research encourages the search for alternatives to GR. 
Usually, gravity can be considered as a fundamental interaction and, from some fundamental principle, the corresponding theory can be constructed. Some examples are $f(R)$ gravity~\cite{Sotiriou:2008rp,odintsov1}, massive gravity~\cite{deRham:2014zqa}, Horndeski~\cite{Horndeski:1974wa}, etc.
{An alternative} approach to understanding the incompatibility of GR with quantum mechanics is to consider gravity as an emergent phenomenon. 
This idea was explored in~\cite{Jac}, by using the Bekenstein-Hawking entropy and the first law of thermodynamics to derive Einstein's equations. 
A resurgence of the subject was ignited by Verlinde's ideas, where he argued that gravity is an entropic force~\cite{Ver,Verlinde2}. This approach is motivated by ideas in holography and uses the
entropy-area relation for black holes. Other proposals make use of the holographic principle, specifically the laws of entanglement, to derive Einstein equations~\cite{vR}.\\ The main ingredient in these formulations is the Bekenstein-Hawking entropy.  
Since these formulations of gravity have an entropic origin, modifications to gravity can be induced by modifying the entropy-area relationship. 
In \cite{Diaz-Saldana:2018ywm}, the authors propose a modified entropy-area relation that, in addition to the Bekenstein-Hawking term, includes a logarithmic (usually related to quantum effects \cite{tkach}), a volumetric and linear terms \footnote{{These contributions to the entropy were obtained from the supersymmetric generalization of the Wheeler-DeWitt (WDW) equation for Schwarzschild black holes.}}.  This new entropy was used to study galactic rotation curves \cite{Diaz-Saldana:2018ywm}, concluding that the volumetric term can account for the anomalous rotation curves and that logarithmic and linear terms have a negligible contribution at the galactic level. In the context of cosmology, in \cite{Shey1,Shey3} the authors considered a
logarithmic correction to the Bekenstein-Hawking entropy and derived the respective Friedmann equations. 
{More recently, it was shown that with the volumetric contribution one can derive a
self-accelerating universe~\cite{Chagoya:2023hjw}; in fact, this model is equivalent to the Dvali-Gabadadze-Porrati (DGP) cosmological model arising in the brane-world scenario~\cite{Deffayet:2001pu}. This correspondence is better understood when one notices that the entropy of DGP has a contribution related to gravity in the bulk, that is, a volumetric contribution~\cite{Sheykhi:2007zp}. 
However, despite its appealing features, the DGP cosmological model faces challenges due to observational inconsistencies and ghost instabilities; also, data from SN Ia, BAO, and the CMB show that the modified Friedmann equation is less compatible with observations than the standard Friedmann equation \cite{Xia_2009}. For this reason, a modified version of the DGP Friedmann equation has been proposed~\cite{Dvali:2003rk}, in which a different power of the Hubble parameter is considered in the Friedmann equation in addition to the usual $H^2$ term.

In this work, following \cite{Cai:2008ys}}, we derive 
a modified Friedmann equation (MFE) 
from
a more general entropy-area relation. In analogy to DGP, we expect the present model to describe a self-accelerating universe; however, we foresee that the additional terms may allow for a better agreement with observations than DGP, while also being compatible with a value of $H_0$ closer to local measurements.

This paper is organized as follows. In Sec.~\ref{mfe} the MFE is obtained starting from a modified entropy-area relation, which is proposed in a generalized manner. In Sec.~\ref{Low} particular cases of the MFE are considered and their behavior is studied in the low-redshift regime.  In Sec.~\ref{contraints}, two particular models are confronted with observational Hubble data and strong lensing systems. These models correspond to considering an entropy that includes the Hawking-Bekenstein, volumetric, and linear terms, and the second case adds the logarithmic term on the entropy-area relationship. 
Lastly, Sec.~\ref{conclusions} is devoted to discussion and final remarks.

\section{Obtaining the Modified Friedmann Equation} \label{mfe}

Let us start by reviewing the derivation of the modified Friedmann equations based on the application of the Clausius relation $\delta Q=T dS$ on the apparent horizon of the FRW universe. For more details on the derivation, see \cite{Cai:2008ys}.
The metric for a spatially-flat FRW universe is expressed as
\begin{equation}
ds^{2}=-dt^{2}+a^{2}(t)\left(dr^2+r^{2}d{\Omega}^{2}\right).
\end{equation}
The apparent horizon is defined by the condition $h^{ab}\partial_{a}\tilde{r}\partial_{b}\tilde{r}=0$, where $\tilde{r}=a(t)r$ and $h^{ab}$ is identified by writing the metric as $ds^{2}=h_{ab}dx^{a}dx^{b} +\tilde{r}^{2}d\Omega^{2}
$. This condition yields $\tilde{r}_{A}=H^{-1}$ for the radius of the apparent horizon.

In order to apply Clausius relation on the apparent horizon, we assume that its temperature is given by $T=1/2\pi \tilde{r}_A$ while its entropy  is given by the following modified entropy-area relationship
\begin{equation}\label{entr2}
S=\frac{A}{4G}+\alpha \ln \frac{A}{4G}+\sum_{j=0}^{N}\sigma_{j}\left(\frac{A}{4G}\right) ^{\frac{1+j}{2}},   \end{equation}
where $A=4\pi \tilde{r}_{A}^2$ is the area of the apparent horizon  and $\alpha,\sigma_{j}$ are free parameters of the model.

This form of entropy is inspired by the results in~\cite{Diaz-Saldana:2018gxx}, where volumetric, linear, and logarithmic modifications were found for the Bekenstein-Hawking entropy. The volumetric dependence is typically related to degrees of
freedom in ordinary quantum field theory, while the linear term has been reported to be an effective contribution due to a self-gravitating gas where $S\sim V^{1/3}$~\cite{deVega:2005gv}. Finally, the logarithmic term seems to be a universal modification to the Bekeinstein-Hawking entropy and has been found in different approaches in the study of black holes~\cite{Obregon:2000zd,Domagala:2004jt,Mukherji:2002de,Sen:2012dw,Chagoya:2023ddb}. The modified entropy considered in~\cite{Diaz-Saldana:2018gxx} corresponds to a special case of Eq.\eqref{entr2} where the series is truncated at $N=2$. Note that the term $j=1$ in the series gives a term proportional to the area; therefore, it is not relevant as far as modifications to the area law are concerned.

The amount of energy $\delta Q$ that crosses the apparent horizon during the time interval $dt$ can be calculated in a straightforward manner~\cite{Hayward:1997jp} by considering the matter content as a perfect fluid, whose energy-momentum tensor is used to obtain $\delta{Q}=A(\rho+P)dt$.
Combining these elements, the Clausius relation yields
\begin{equation}\label{PreMFE}
\frac{4\pi G}{3}\dot{\rho}=\left[1+\frac{4G\alpha}{A}+ 
\sum_{j=0}^{N}\sigma_{j}(4G)^{\frac{1-j}{2}}\left(\frac{1+j}{2}\right)A^{\frac{j-1}{2}}\right]\dot{H}
\end{equation}
where the continuity equation $\dot{\rho}+3 H(\rho+P)=0$ has been used. Finally, plugging $A=4\pi H^{-2}$, Eq.\eqref{PreMFE} can be integrated, yielding the MFE
\begin{equation}\label{fried}
\frac{8\pi G}{3}\rho= H^2+\frac{G\alpha}{2\pi}H^4+\sum_{j=0}^{N}\sigma_{j}\left(\frac{\pi}{G}\right)^{\frac{j-1}{2}}\frac{1+j}{3-j}H^{3-j},
\end{equation}
with $\sigma_{3}=0$. Let us rewrite the MFE as follows
\begin{equation}\label{friedomegas}
(1+z)^{3(1+\omega)}\Omega_{0m}= E(z)^2+\Omega_{0\alpha}E(z)^4+\sum_{j=0}^{N}\Omega_{0\sigma_{j}}E(z)^{3-j},
\end{equation}
where $E(z)=H(z)/H_{0}$ and we have introduced the density parameters at present time $\Omega_{0\sigma_{j}}$, and $\Omega_{0\alpha}$, associated to the free parameters $\alpha$ and $\sigma_{j}$, respectively. These are defined as
\begin{equation}\label{def:omegas}
\Omega_{0\sigma_{j}}=   \sigma_{j}\left(\frac{\pi}{G}\right)^{\frac{j-1}{2}}\frac{1+j}{3-j}   H_{0}^{1-j},\quad  \Omega_{0\alpha}= \frac{G H_{0}^{2}}{2\pi}\alpha,
\end{equation}
while the matter density parameter at present time, $\Omega_{0m}$ is defined in the usual manner as
 \begin{equation}
\Omega_{0m}=\frac{8\pi G\rho_0}{3H_{0}^{2}}. 
\end{equation}
We also consider a barotropic fluid with an equation of state that implies $\rho(z)=\rho_{0}(1+z)^{3(1+\omega)}$.  In this paper, we restrict our analysis to $\omega=0$, which corresponds to dust, while the case $\omega=1$ is also interesting and will be discussed in the final section.

Although the modification terms in the series for $N > 2$ might not have a foundation in a certain gravitational theory, they can be motivated as a result of an effective cosmological theory, and thus we include them to express the entropy-area relation in a generalized manner. For example, if a certain power of \( H \) appears in the MFE of some cosmological model, it is possible to determine the corresponding power of \( A \) in the entropic modification that would give rise to such term. For instance, in a recent work \cite{Chen:2024ufj}, the authors consider a dark energy model whose Friedmann equation includes a term proportional to \( H^{-2} \), which can be mapped to an entropic modification term proportional to \(A^{3} \). This corresponds to the term \( j=5 \) in the series in Eq.\eqref{entr2}. Therefore, the MFE considered in the mentioned work is a particular case of Eq.\eqref{fried} with $N=5$, $\alpha=0$, and $\sigma_{0}= \sigma_{1}=\sigma_{2}=\sigma_{4}=0$. Another example follows from the MFE of the DGP model, which includes a modification term proportional to $H$, this corresponds to a special case of Eq.\eqref{fried} truncated at $N=2$ with $\alpha=\sigma_{0}=\sigma_{1}=0$, however, the entropic modification that gives rise to such term in the MFE is the one proportional to $A^{3/2}$ which, as pointed out before, has been obtained in the literature. Similarly, in~\cite{Lima:2013dmf}, the authors propose a model based on a
dynamical vacuum energy density, which introduces even powers of the Hubble parameter. These terms might be assigned to entropic modifications of negative powers of the area, which are not considered in this work. Another motivation to consider the $N>2$ entropic modifications in Eq.\eqref{entr2} may arise when aiming to adjust the behavior of \( H(z) \) at different redshifts: the inclusion of additional terms in the entropy introduces new terms to the MFE which become relevant at different stages in the evolution of the Universe as the area increases. This will be further detailed in the last section.  
\section{Low-Redshift behavior and Entropic Contributions}\label{Low}

\begin{figure}[h!]
\begin{center}
\includegraphics[scale=.18]{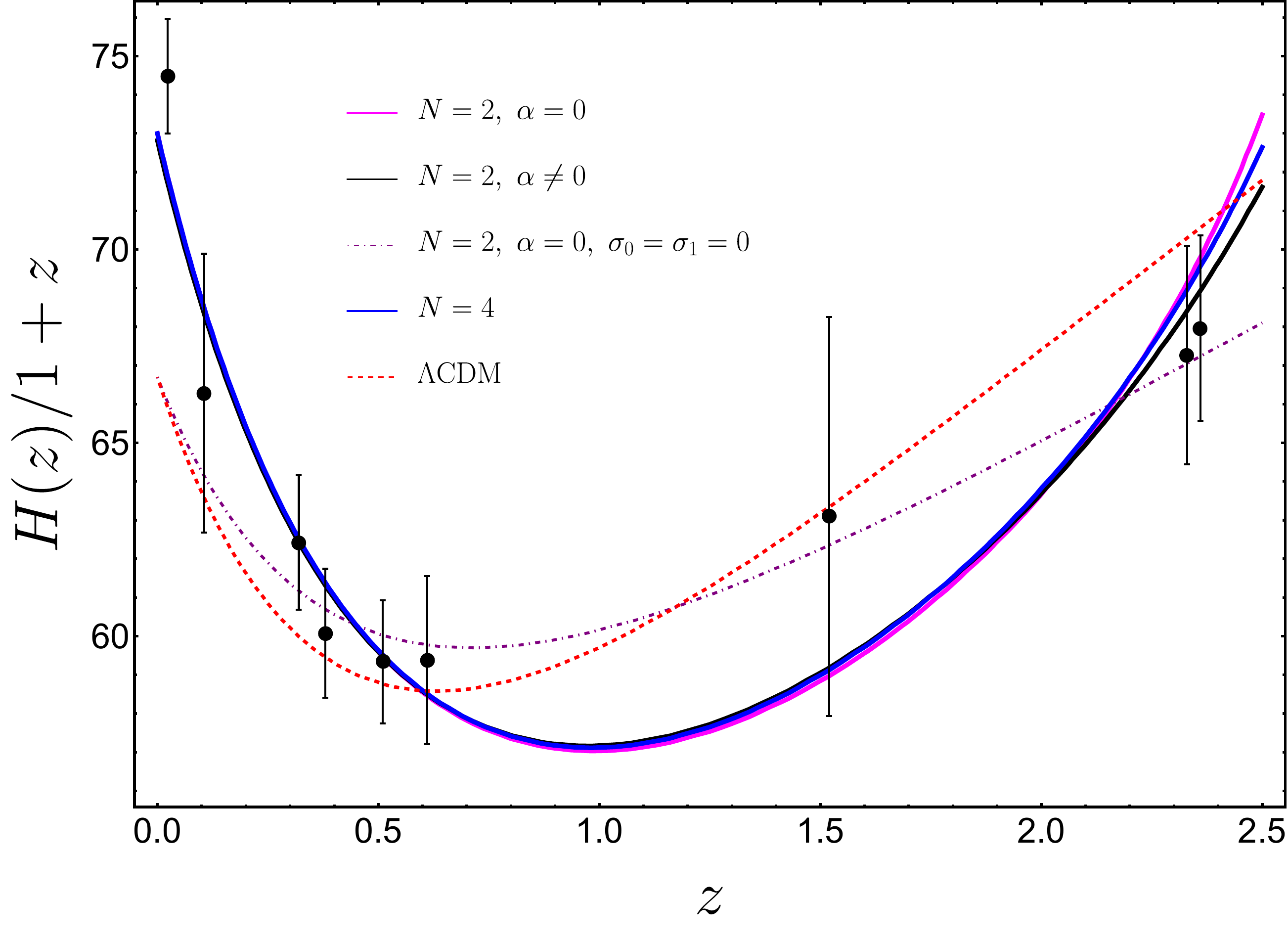}
\caption{Low-redshift behavior of $H(z)/(1+z)$ for different MFEs arising from Eq.\eqref{friedomegas}, compared to the $\Lambda$CDM model. Data points shown are representative of the compilation presented in \cite{Magana:2017nfs}.} \label{hprofiles}
\end{center}
\end{figure}

The MFE obtained from Eq.\eqref{friedomegas} is able to reproduce the expected behavior of $H(z)$ in the low redshift regime. In Fig.~\ref{hprofiles}, we show representative curves of $H(z)/(1+z)$ for three different MFEs. These curves correspond to $N=2$ with $\alpha=0$ and $\alpha \neq 0$, and $N=4$, respectively. In addition, we include the MFE of the DGP model and $\Lambda$CDM. Also, some observational points of $H(z)$ are included in order to visualize the behavior of the models (details on observational data will be presented in section~\ref{contraints}). It can be seen that the solid curves exhibit a similar behavior for certain values of the free parameters. Moreover, even with a high value of $H_0$, these curves demonstrate a good fit to observations. In contrast, the DGP model and $\Lambda$CDM require a lower value of $H_0$ to fit the observational data.
For this reason, if we aim to focus on the redshift region, where the majority of available observational data are concentrated, we are motivated to truncate the series in Eq.\eqref{fried} to \( N=2 \). For this particular case, the resulting MFE is 
\begin{align}
(1+z)^{3}\Omega_{0 m}&=
E^2-\Omega_{0\epsilon}E-\Omega_{0\beta}E^3 + \Omega_{0\alpha}E^4,\label{eq:quartic}
\end{align}
 where the parameters $\Omega_{0\beta}, \Omega_{0\epsilon}$ are obtained from Eq.\eqref{def:omegas} as
\begin{equation}
    \Omega_{0\beta}=  H_{0} \sqrt{\frac{G}{9\pi}}\beta, \  \Omega_{0\epsilon}=\frac{3}{H_0}\sqrt{\frac{\pi}{G}}\epsilon,
\end{equation}
and we have defined $\beta=-\sigma_{0} \ \epsilon=-\sigma_{2} $. The density parameters in Eq.\eqref{eq:quartic} obey the constraint
\begin{equation}
\Omega_{0m} +\Omega_{0\epsilon}
+\Omega_{0\beta}- \Omega_{0\alpha} = 1,\label{eq:z0cond}
\end{equation}
which follows by evaluating Eq.\eqref{eq:quartic} at $z=0$.
\section{Observational constraints}\label{contraints}
In order to assess the observational viability of the entropic models, we constrain their free parameters with observational data from cosmic chronometers (CC), baryon acoustic oscillations (BAO), and strong lensing systems (SLS). We present the methodology used for each data set.

\subsection{Hubble parameter measurements}
We use the original Hubble data (OHD)  compiled in \cite{Magana:2017nfs}, which include 31 CC measurements in the redshift range $0.07 < z < 1.965$ and 20 additional BAO data points spanning the redshift interval $0.24 < z < 2.36$.  
In order to constrain the free parameters of the model, we employ the following chi-square function,
\begin{equation}
\chi_{cc}^2 = \sum_{i=1}^{N} \frac{ \left[ H(z_{i}) -H_{obs}(z_{i})\right]^2 }{ \sigma_{H_i}^{2} }, \label{chiCC}
\end{equation}
where the number of data points is denoted by $N$, 
$H_{obs}(z_{i})$ represents the measured value at $z_{i}$, $\sigma_{H_i}$ is the error in each measurement, and $H(z_{i})$ indicates the theoretical value for a given model\footnote{We can also replace $H(z_{i})$ by $H(q_0,j_0,s_0,l_0)$ to constrain cosmographic parameters using Eq.\eqref{chiCC}}.
\subsection{SLS measurements}
We use a fiducial sample from the latest SLS compilation provided in \cite{Amante:2019xao}. This sample includes 143 observations of early-type galaxies acting as gravitational lenses and provides four observed properties: spectroscopically determined stellar velocity dispersion
$\sigma$, the Einstein radius $\theta_E$, the lens redshift $z_l$, and the source redshift $z_s$. These measurements can be utilized as a tool for testing cosmological models through the following chi-square function
\begin{equation}
\chi_{sl}^2(\Theta) = \sum_{i=1}^{N_{SLS}} \frac{ \left[ D^{th}\left(z_{l}, z_{s}; \Theta \right)  -D^{obs}(\theta_{E},\sigma^2)\right]^2 }{ (\delta D^{\rm{obs}})^2},
\label{eq:chisquareSL}
\end{equation}
where $N_{SLS}$ accounts for the number of strong lensing systems,  $D^{th}$ is the theoretical distance
ratio $D \equiv D_{ls}/D_{s}$ between the angular diameter distance from the lens to the source ($D_{ls}$) and 
from the observer to the source ($D_{s}$), and $\delta D^{\rm{obs}}$ is the standard error propagation of the observational lens equation ($D^{obs}$), which in this case is defined through the Einstein radius of the singular isothermal sphere (SIS) model as
\begin{equation}
D^{obs} = \frac{c^2 \theta_{E}}{4 \pi \sigma^2}, \label{Dlens}
\end{equation} 
where $c$ is the speed of light.
\subsection{Methodology}
As already mentioned, if we consider the area term and the volumetric term in the entropy, the model corresponds to the DGP model. Adding the length term on the entropy gives a Friedmann equation that has a cubic term on the Hubble parameter. If we also include the logarithmic term, the Hubble parameter on the Friedmann equation is of quartic order. For this reason, the modified Friedmann equations are labeled as MFE3 and MFE4. From MFE3 it is possible to obtain a tractable analytic expression for $E(z)$. 
Fig.~\ref{hprofiles}, shows that this truncation is justified for $z\leq 2.5$.
The resulting equation
is such that its discriminant at $z=0$ is always positive (assuming $\Omega_{0m}<1$), thus, in this limit the MFE3 has three distinct real solutions. The solution that behaves appropriately at $z=0$ is 
\begin{equation}
E(z) = \xi C -\frac{p}{3\xi C} +\frac{1}{3\Omega_{0\beta}},\label{eq:ez3}
\end{equation}
where
\begin{align}
    p& = \frac{3 \Omega _{{0\beta}} \Omega _{{0\epsilon}}-1}{3 \Omega _{{0\beta }}^2},\label{eq:p}\\
    q& = \frac{9 \Omega _{{0\beta }} \left[3 (z+1)^3 \Omega _{\text{0$\beta $}} \Omega _{{0m}}+\Omega _{{0\epsilon}}\right]-2}{27 \Omega _{{0\beta}}^3}, \\
C & = \left({\sqrt{\frac{p^3}{27}+\frac{q^2}{4}}-\frac{q}{2}}\right)^{1/3},
\end{align}
and $\xi = \left(-1-i \sqrt{3}\right)/2$. We remark that despite the appearance of $\xi$, this root is real.  
\begin{figure}
\includegraphics[scale=.235]{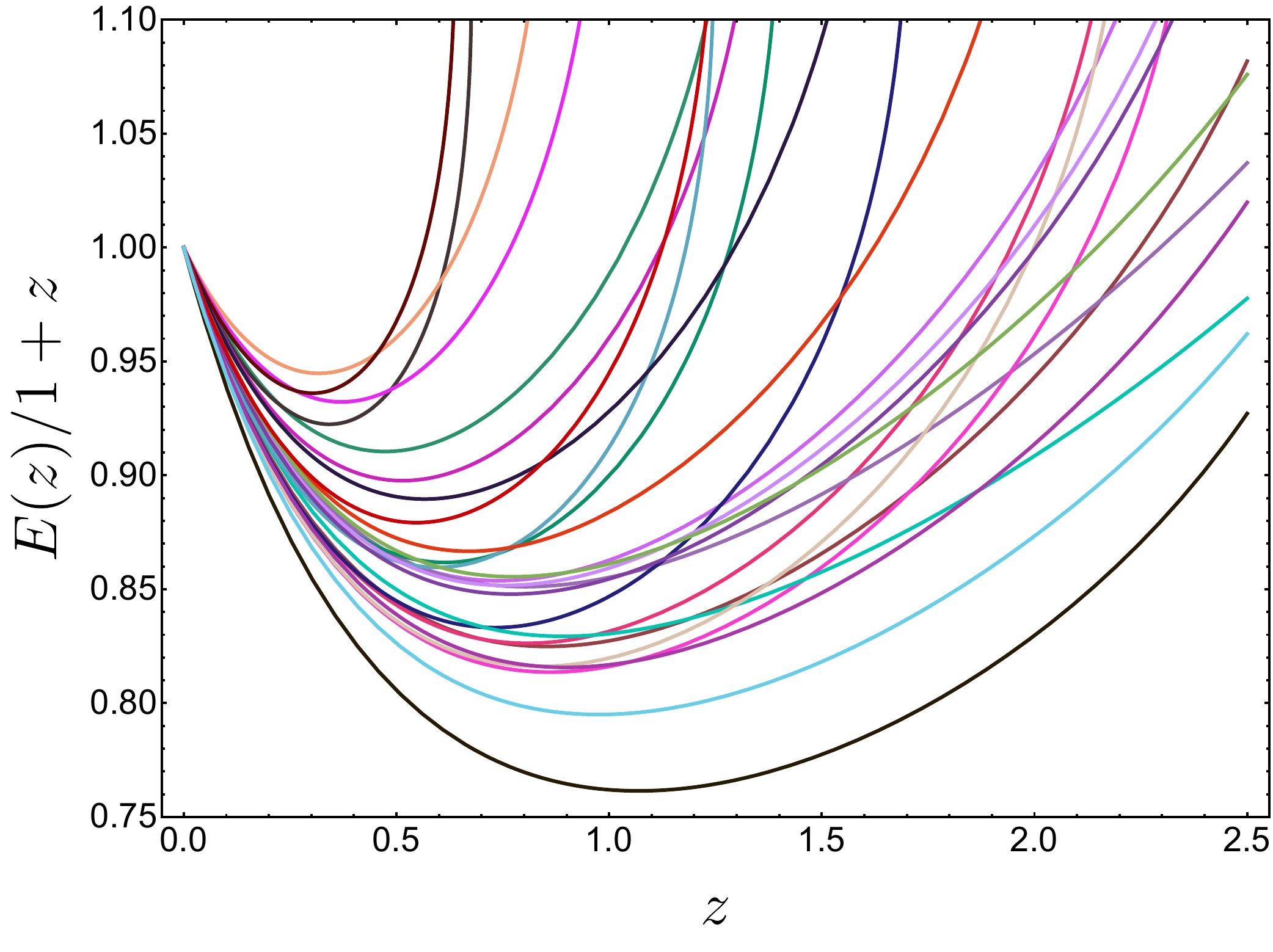}
    \caption{Profiles of $E(z)$, Eq.\eqref{eq:ez3}, for random values of $\Omega_{0m}\in[0.08,0.15]$ and $\Omega_{0\epsilon}\in[0.6,0.8]$, and $\Omega_{0\beta} = 1 -\Omega_{0m}-\Omega_{0\epsilon}$.}
    \label{fig:hz3}
\end{figure} 
Several profiles of $E(z)$ are shown in Fig.~(\ref{fig:hz3}), with random values of the density parameters at present time -- only constrained by Eq.\eqref{eq:z0cond} -- with the aim of showing that small variations of these parameters lead to diverse evolutions of $E(z)$. This observation is relevant for selecting the priors of the statistical analysis presented below.

\begin{figure}[H]
    \centering    \includegraphics[width=0.45\textwidth]{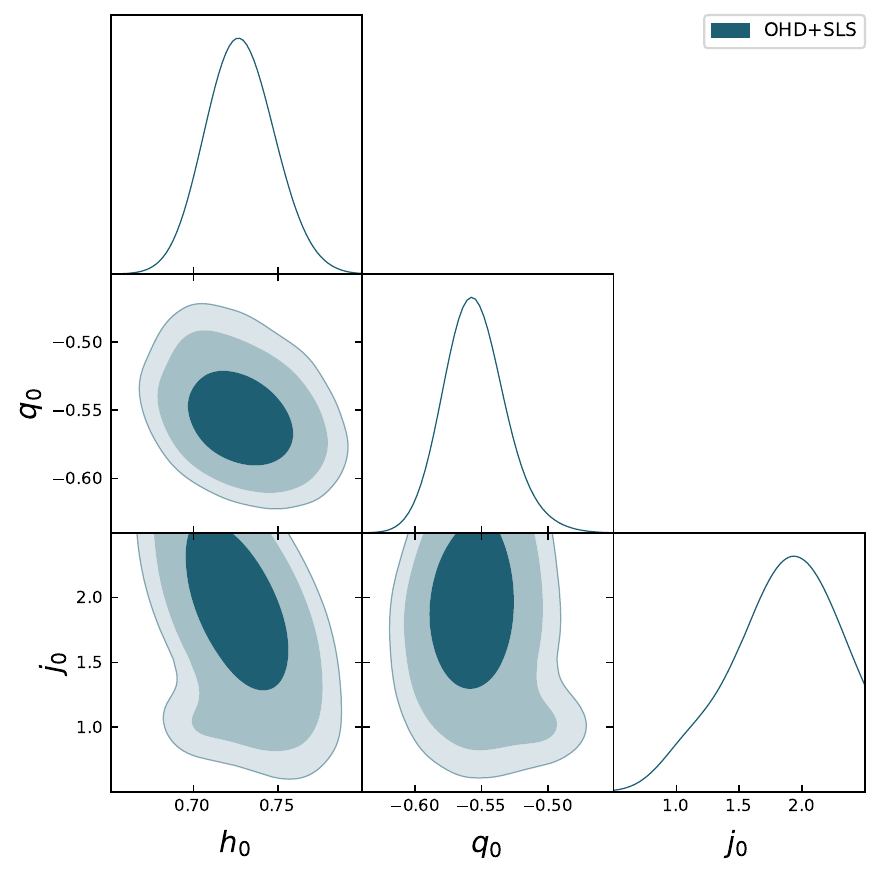}
    \caption{Cosmographic contours from a joint analysis OHD+SLS. The best fit parameters are $h_0 = 0.727^{+0.020}_{-0.019} $, $q_0=-0.557^{+0.022}_{-0.021}$, $j_0=1.895^{+0.346}_{-0.407}$.}
    \label{fig:enter-label}
\end{figure}

\begin{figure}[H]
    \includegraphics[scale=0.6]{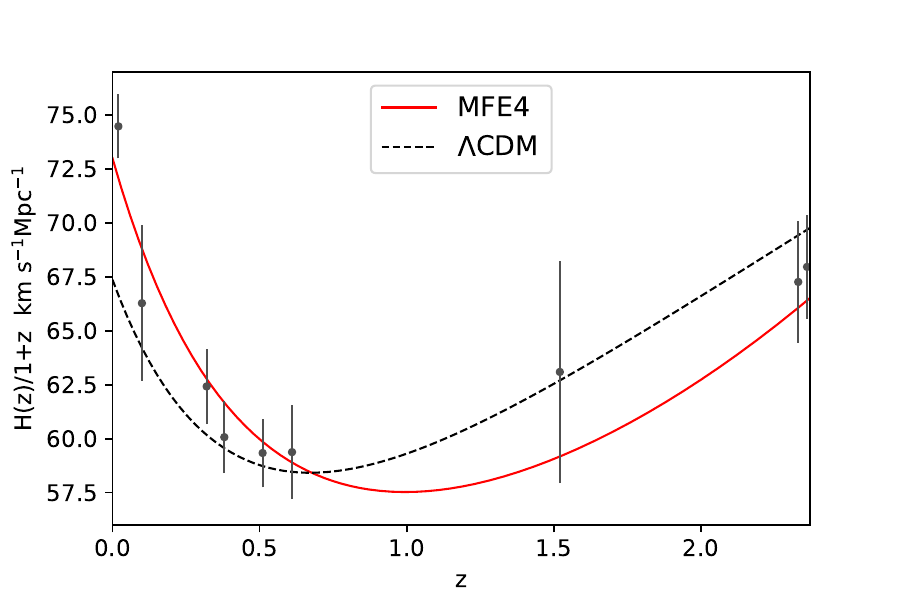}
\caption{Hubble parameter reconstruction for the MFE4 and $\Lambda$CDM models. For the MFE4, we used the cosmological parameters reported in Table~\ref{tab:ocparnueva}. For $\Lambda$CDM we assume the parameters obtained from Planck measurements~\cite{Planck:2018vyg}. Data points shown are representative of the compilation presented in \cite{Magana:2017nfs}.}
    \label{Hzmodels}
\end{figure}

In order to constrain the parameters of the MFE3, we  employ a Markov Chain Monte Carlo analysis (MCMC) through \emph{emcee}~\cite{Foreman-Mackey:2012any} using both, the $H(z)$ and SLS measurements, as a joint analysis i.e. $
\chi_{{joint}}^2 = \chi_{{cc}}^2 + \chi_{{sl}}^2 
$. 
Using the analytical expressions derived from the MFE3, the MCMC process fails to identify a region of maximum probability unless the priors are extremely narrow. We attribute this to the high variability of $E(z)$ displayed in Fig.(\ref{fig:hz3}), which affects the convergence of MCMC. 
As an alternative approach, we constrain the model indirectly by using the following $y$-redshift cosmographic expansions\footnote{The inclusion of the $y$-redshift parametrization provides enhanced constraints on the cosmographic parameters as has been shown previously in \cite{Lizardo:2020wxw,Amante:2024ett,Zhang:2023eup}}~\cite{Cattoen:2007sk}
\begin{align}
    H(y)  = &   H_0 \Big[ 1 + (1+q_0)y +\frac{1}{2}(2+2q_0-q_0^2 + j_0)y^2 \nonumber \\
     &+ \frac{1}{6}(6+6q_0-3q_0^2 + 3q_0^3 -4q_0j_0 + 3j_0 -s_0)y^3 \nonumber \\
    &+\mathcal{O}(y^4) \Big] \label{Hubbley},
    \end{align}
with $y(z)=z(1+z)^{-1}$ and $q_0, j_0$ are the cosmographic parameters, and
\begin{align}
D^{th}\left(y_{l}, y_{s}; \Theta \right)
& \approx \frac{y_s-y_l}{y_s-y_sy_l}-\frac{(1+q_0)(y_s-y_l)y_l}{2(y_s(-1+y_l)^2)} + \mathcal{O}[y^2],\label{Dthy}
\end{align}
where $y_{l}$ and $y_s$ are related to the redshift of the lens and the redshift of the source, respectively. Relating the parameters of the model with the cosmographic ones by Taylor expanding Eq.\eqref{eq:ez3}, the cosmographic parameters are
constrained with the same data sets described above. The corresponding confidence contours for the cosmographic parameters are shown in Fig.~(\ref{fig:enter-label}). {The obtained parameter $h_0$ shows strong consistency within  0.15$\sigma$  with the value reported by Riess et al. \citep{Riess:2021jrx}, while $q_0$ remains reasonably consistent up to 1.46$\sigma$ with the same work. Although $j_0$ does not have a direct comparison with this study, it is observed to deviate approximately 2$\sigma$ from the standard model, where $j_0=1$.} Based on previous work \cite{Amante:2024ett}, the difference between the best fit parameters of a modified gravity model and those inferred from cosmography may be up to $3\sigma$. By assumption, the best-fit parameters of the MFE3 are close to those of the MFE truncated at higher orders when using observations at low redshift. Therefore, in the following, we consider the MFE4 and refine the search for the best-fit parameters in a region that is within the $3\sigma$ cosmographic confidence levels and such that the solution to the cubic equation remains real up to $z=2.5$. 
Specifically, we reduce the interval of interest to
$\Omega_{0m}\in[0.085,0.12], \Omega_{0\epsilon}\in[0.76,0.83]$, while $\Omega_{0\alpha} = 0.008$ is arbitrarily fixed to a small value and $\Omega_{0\beta} = 1 - \Omega_{0m}-\Omega_{0\epsilon}+\Omega_{0\alpha}$. Since the solution for $H(z)$ for the MFE4 is not analytically tractable, we numerically 
generate families of curves for $H(z)$ by varying the model parameters. Each curve is tested against each data set through a joint analysis $
\chi_{joint}^2 = \chi_{cc}^2 + \chi_{sl}^2, \label{chiJ}
$ and from this analysis we determine the curve $H(z)$ that leads to the minimum $\chi_{joint}^2$. The values of the best-fit parameters are shown in Table~(\ref{tab:ocparnueva}). These best-fit parameters provide insight into the compatibility of the quartic model with observational data. In addition, in Fig.~(\ref{Hzmodels}) we display $H(z)/(1+z)$ for the standard cosmological model for comparison.
\begin{table}
\caption{Best-fit parameters for the MFE4 derived from OHD+SLS data.}
\centering
{
\begin{tabular}{|ccccccc|}
\hline
\multicolumn{7}{|c|}{MFE4 model (OHD + SLS)}\\ \hline
 $h_0$ & $\Omega_{0m}$  & $\Omega_{0 \epsilon}$ &$\Omega_{0 \beta }$ &$\Omega_{0 \alpha}$ & $\chi^{2}_{min}$& $\chi^{2}_{red}$ \\

0.727 &  $0.095$ &$0.770$ &0.143 & 0.008    & 286.455  & 1.508 \\
\hline
\end{tabular}}
\label{tab:ocparnueva}
\end{table}

\section{Discussion and final remarks}\label{conclusions}
We proposed a modified entropy-area relation that generalizes other modifications that have been reported in the literature. After studying the consequences of our general relation for a cosmological scenario, we find that for redshift $z\lesssim 2.5$ a combination of well-known entropies gives good agreement with observations of the evolution of the Hubble parameter. 

In agreement with previous studies, our results show that the entropic modifications to the Friedmann equation are able to mimic the effects of the dark components (dark energy or dark matter). In fact, for $z<1$ the quartic model specified in Table~\ref{tab:ocparnueva} can be fitted to a $\Lambda$CDM-like model containing only matter and cosmological constant with
densities $\Omega_{0m} = 0.21$ and $\Omega_\Lambda = 0.79$
and $H_0 = 73.33$ Km s${}^{-1}$ Mpc${}^{-1}$. These can be considered as effective matter and cosmological constant density parameters of the quartic model at low redshift.
Regarding the statistical significance of our results, using SLS and OHD data, we find that our MFE4 provides competitive constraints, producing values of $\chi^{2}_{red}$ (see Table \ref{tab:ocparnueva}) closely comparable to those obtained in the standard model $\chi^{2}_{red}=1.552$ using the same data set. It is expected that incorporating entropic corrections into cosmological dynamics could potentially account for the Universe's accelerated expansion without requiring a dark energy component, such as the cosmological constant employed in the prevailing standard model of cosmology. Based on our constraints, the $\Omega_{0 \epsilon}$ parameter appears to play the role of the agent responsible for the current acceleration of the Universe, slightly exceeding the contribution of the cosmological constant in the standard paradigm at present time. 
It should be noted that the models employed in this work offer a potential resolution to the Hubble tension, allowing $H_0$ values consistent with those reported by Riess et al. \citep{Riess:2021jrx} based on local observations. However, further investigation is required to assess the model's viability in light of early Universe observations. {In contrast, the value of $H_0 = 69.422^{+0.861}_{-0.858}$ obtained by constraining the standard model using the same data set differs by $1.8$ $\sigma$ from the value inferred by the SH0ES program. The remaining parameters of the $\Lambda$CDM model constrained by OHD+SLS data are $\Omega_{0CDM} = 0.230^{+0.012}_{-0.012}$ and $\Omega_{0b} = 0.049^{+0.001}_{-0.001}$.}

The different entropic contributions influence the dynamics of the Universe at different epochs. Modifications involving positive powers of the area only play a significant role at low redshifts. For instance, the DGP model, which incorporates the volumetric contribution, exhibits the characteristics of a self-accelerating universe; however, when the linear contribution is added, the behavior of $H(z)$ near $z = 0$ is adjusted, improving the agreement with the observational data. Meanwhile, the logarithmic contribution becomes particularly relevant at large redshifts, influencing the evolution at large $z$. 

\begin{figure}
\includegraphics[width=0.5\textwidth]{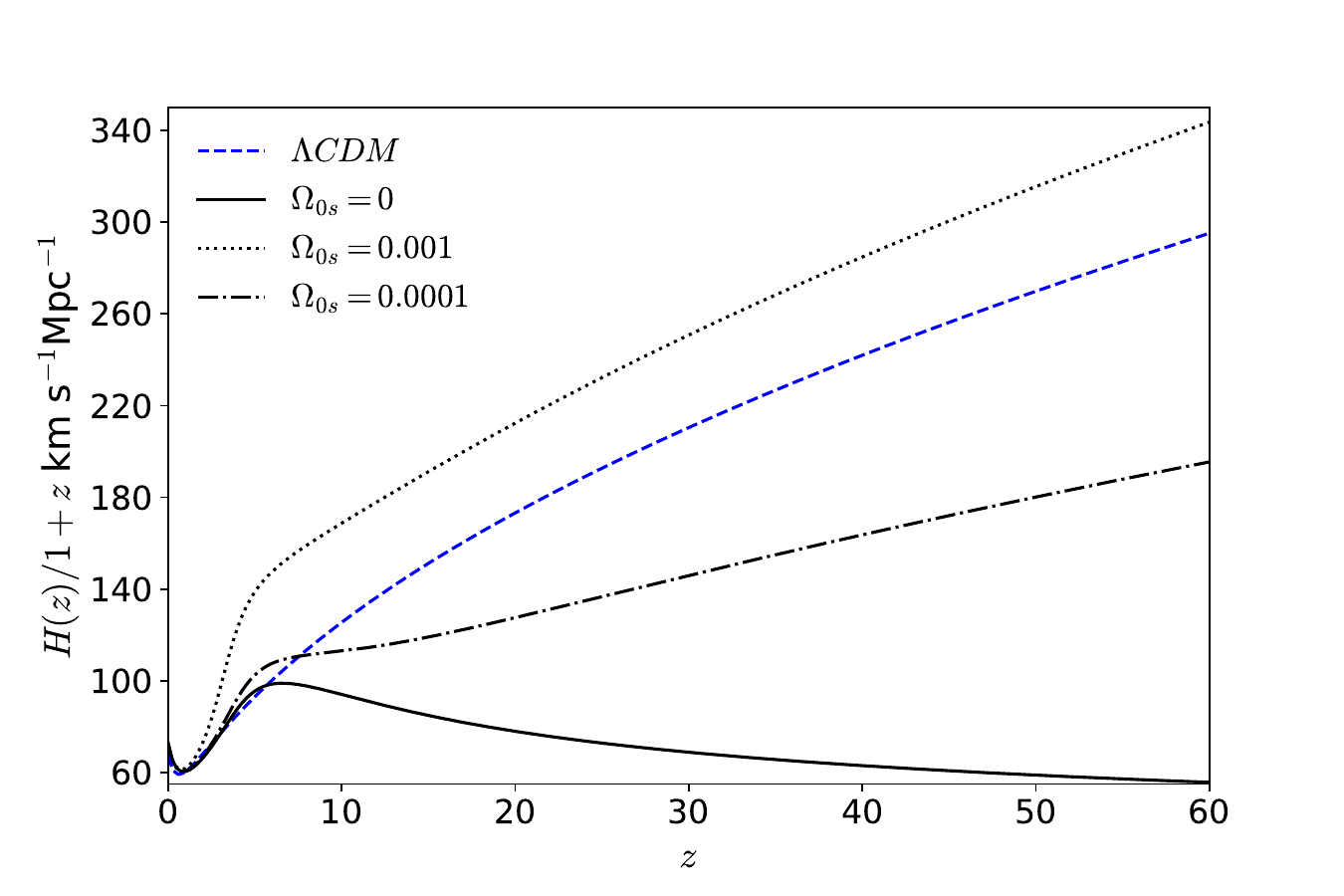}
\caption{Comparison of different cosmological models. A simplified $\Lambda$CDM with $H_0=67.34$\ Km\,s${}^{-1}$ Mpc${}^{-1}$, $\Omega_{0m}=0.315$ and $\Omega_\Lambda = 1 - \Omega_{0m}$ is shown with dashed line. The other lines correspond to solutions of $\mathcal F_E+\mathcal F_L+\mathcal F_{GE}$ with $\Omega_{0m} = 0.120, \Omega_{0\beta}=0.124$, $\Omega_{0\epsilon}=0.777$, different amounts of stiff matter $\Omega_{0s}$, and the remaining density parameter, associated to $H^{-1}$ in the Friedmann equation, dictated by the constraint that the sum of all density parameters equals one.}\label{fig:combined}
\end{figure}

{Now we turn our attention to higher powers of  $A$  on the entropy.} The modified entropy can be divided as follows,  the logarithmic term $S_L$, $S_E$ contains the linear, surface and volumetric terms, and the generalized entropic terms $S_{GE}$  of order $A^{5/2}$ and higher,
with the entropy given as the sum $S=S_L+S_E+S_{GE}$, and the corresponding MFE given by
\begin{equation}\frac{8\pi G }{3}\rho = \mathcal F(H),\label{eq:mfe}\end{equation}
where the function $\mathcal F(H) =\mathcal F_L(H) + \mathcal F_E(H) + \mathcal F_{GE}(H)$ also splits into contributions from the different entropic terms,
\begin{align}
\mathcal F_L(H) & = \frac{G\alpha}{2\pi}H^4 ,\quad\mathcal F_E(H) =  H^2- \sqrt{\frac{9\pi}{G}}\epsilon H - \sqrt{\frac{G}{9\pi}}\beta H^{3}, \nonumber\\
\mathcal F_{GE}(H) & =5 \gamma \left(\frac{\pi}{G}\right)^{3/2} H^{-1} + \mathcal{O}(H^{-3}),
\end{align}
with $\sigma_0= -\beta$, $\sigma_{1}= -\epsilon$, and $\sigma_{4}=\gamma$.
It is natural to question what the relevance of these different contributions is for the evolution of the Hubble parameter. Since one motivation to work with modified Friedmann equations is that an effective dark energy emerges from these modifications, we start by considering that the left-hand side of Eq.\eqref{eq:mfe} contains only dust, i.e. $\rho(z)\sim (1+z)^3$. Then we analyze the following scenarios:
\begin{description}
\item[Dust and $\mathcal F_E$] In this case $H(z)$ generically becomes complex for $z<2.5$, so this cannot be considered a complete model.
\item[Dust and $\mathcal F_E+\mathcal F_L+\mathcal F_{GE}$] Including $\mathcal F_L$ and the first term of $\mathcal F_{GE}$ ($\sim H^{-1}$) alleviates the problem of $H(z)$ becoming complex; however, it typically leads to a high redshift behavior very different from $\Lambda$CDM, as exemplified by the solid curve in Fig.~(\ref{fig:combined}). {This behavior is dominated by $\mathcal F_L$.}
\end{description}
{
This suggests that entropic corrections with $N > 2$  are not determinant for the dynamics of the early Universe, although they may contribute to a refined understanding of cosmic evolution. In order to obtain early Universe dynamics that resemble $\Lambda$CDM, one approach is to include other matter contents, in particular a fluid whose density decays faster than the density of radiation, for instance, a stiff fluid -- a type of matter that is not part of the standard cosmological model but appears in several alternative models (e.g. scalar field dark matter~\cite{Li:2016mmc} and $f(R)$ gravity~\cite{Odintsov:2017cfr}).
With the addition of this fluid it is possible to obtain Hubble parameter profiles similar to those of $\Lambda$CDM at large redshift, without spoiling the agreement with low-redshift observations, as shown by the curves with $\Omega_{0s} \neq 0$ in Fig.~(\ref{fig:combined}), where $\Omega_{0s}$ is the density parameter of the stiff component. The role of the $N>2$ entropic contributions, which might be thought of as an effective description of the gravitational dynamics, is to fine-tune such behavior, and they could be constrained with high redshift observations. In this respect, the next step would be to consider future estimations of the Hubble parameter coming from quasars that are expected to be observed by the Dark Energy Spectroscopic Instrument~\cite{DESI:2024lzq}. One could also consider gamma-ray bursts at even redshift up to $z\sim 9$; however, one needs to be careful with the model dependence of these results~\cite{Luongo:2021pjs}. 
}

Going to even higher redshift, one should notice that in addition to the logarithmic term, one can also consider the negative powers of $A$ in the entropy. These terms have been discussed in connection to quantum gravity. Moreover, it is conjectured that to include quantum gravity effects \cite{tkach,Kaul:2000kf,Meissner:2004ju}, the semiclassical Hawking-Bekenstein entropy is generalized as follows 
\begin{equation}
S= \frac{A}{4G}+\alpha \ln \frac{A}{4G}+ \mathcal{O}\left(\frac{1}{A} \right).
\end{equation}

Modifications involving negative powers of the area give higher powers of the Hubble parameter in the Friedmann equations and therefore may become relevant at large redshifts. Since logarithmic corrections to the entropy are related to quantum effects, it is plausible that including negative powers of $A$ in the entropy might provide valuable insights into the behavior of the Universe during earlier epochs. This and other ideas are being investigated and will be presented elsewhere.
\section*{Acknowledgments}
This work is supported by CONAHCYT grants DCF-320821, 257919, 258982, and CIIC 034-2024. {\bf IDS} and {\bf MHA} are supported by the program ``Estancias Postdoctorales por México'', CONAHCyT. {\bf JCLD} is supported by UAZ-2024-39113 grant.

\bibliographystyle{unsrt}
\bibliography{references}






\end{document}